\documentclass{aa}  
\usepackage[english]{babel}
\usepackage{txfonts}
\usepackage{graphicx}
\usepackage[utf8]{inputenc}
\usepackage{amsmath}
\usepackage{xcolor}
\definecolor{indigo(english)}{RGB}{80, 40, 200}

\title{The slingshot effect as a probe of transverse motions of galaxies}
\author{R. Hagala \inst{1}
	\and 
    C. Llinares \inst{2}
    \and
    D. F. Mota \inst{1}}

   \institute{Institute of Theoretical Astrophysics, University of Oslo, PO Box 1029 Blindern, 0315 Oslo, Norway
              \\
              \email{robert.hagala@astro.uio.no}
         \and
Institute of Cosmology and Gravitation, University of Portsmouth, Dennis Sciama Building, Portsmouth PO1 3FX, United Kingdom
             }
\date{}
\abstract{There are currently no reliable methods to measure transverse velocities of galaxies. This is an important piece of information that is lacking in galaxy catalogues, and could allow us to probe the physics of structure formation as well as testing the underlying theory of gravity. The slingshot effect---a special case of the Integrated Sachs--Wolfe effect---is expected to create dipole signals in the temperature fluctuations of the Cosmic Microwave Background Radiation (CMB). This effect creates a hot spot behind and a cold spot in front of moving massive objects. The dipole signal created by the slingshot effect can be used to measure transverse velocities, but because the signal is expected to be weak, the effect has not been measured yet.}
{The aim is to show that the slingshot effect can be measured by stacking the signals of galaxies falling into a collapsing cluster. {Furthermore, we evaluate if the effect can probe modified gravity.}}
{We use data from a simulated galaxy catalogue {(MultiDark Planck 2)} to mimic observations. We identify a $10^{15} M_\odot$ cluster, and make maps of the slingshot effect for photons passing near 8438 infalling galaxies. {To emulate instrument noise, we add uncorrelated Gaussian noise to each map. We assume the average velocity is directed towards the centre of the cluster; The maps are rotated according to the expected direction of motion. This assures that the dipole signal will add up constructively when stacking the maps.} We compare the stacked maps to a dipole stencil to determine the quality of the signal. We also evaluate the probability to fit the stencil in the absence of the slingshot signal.}
{Each galaxy gives a signal of around $\Delta T/T \approx 10^{-9}$, while the precision of CMB experiments of today are $\Delta T/T \approx 4 \times 10^{-6}$. By stacking around 10 000 galaxies and performing a stencil fit, the slingshot signal can be over the detectable threshold with experiments of today. However, due to the difficulty of distinguishing an actual signal from false positives, future CMB experiments must be used to be certain of the strength of the {observed} signal.}
{}

\begin{document}

\maketitle

\section{Introduction}
By precisely measuring the positions and velocities of galaxies, one can use them as tracers for mapping the underlying matter distribution of the large-scale structure of the Universe. Furthermore, in systems where the matter distribution is known from other methods, for instance from gravitational lensing, precise catalogues of the positions and velocities of galaxies can be used as a consistency check to test our models of gravity and structure formation. 
While the radial velocity with respect to the earth is measurable through the Doppler effect, transverse velocities of galaxies are more challenging to measure. The only reliable method to estimate transverse velocities of an object directly is through detecting a change in position relative to the background between two observations, a so-called proper motion.
The recent data release of Gaia \citep{GaiaCollaboration2018GaiaProperties} presented the proper motions of over a billion stars in our galaxy, which were found using this method.
However, this option becomes unfeasible at scales larger than our galaxy because distant galaxies will need many years to move far enough for the motion to be resolved. 

{
When a galaxy is affected by an external force, like gravity, the resulting acceleration will first change the velocity of the galaxy before the position of the galaxy changes significantly. Better knowledge of the peculiar velocities of observed galaxies will help us compare our models of gravity on large scales with the forces acting on those galaxies. This can be particularly useful when studying theories of modified gravity, but can also be used to measure the amount and distribution of dark matter in the context of $\Lambda$CDM. Velocities can be used as a consistency check for our models of structure formation. The best models predict the matter distribution very well, but the velocity field of the same models are not usually compared to observations \citep{Stebbins2006MeasuringLSS}.
Furthermore, measurements of tangential velocities can break the degeneracy between expansion velocity and peculiar velocity, which is a problem when using only the redshift to measure velocities. }

The slingshot effect was first mentioned by \cite{Birkinshaw1983AGalaxies}, and is a promising probe for measuring transverse motions. It is a special case of the Integrated Sachs--Wolfe (ISW) effect \citep{Sachs1967PerturbationsBackground}. The ISW contribution to the Cosmic Microwave Background (CMB) is due to the evolution of a gravitational potential while photons are passing through it. The most known result of the late-time ISW is what is known as the Rees--Sciama effect \citep{Rees1968Large-scaleUniverse};  As structure collapses under gravity---or expands with the Hubble flow---while a photon passes through, the change in the gravitational potential of the structure will affect the photon energy. The Rees--Sciama effect results in an overall increase or a decrease in the measured CMB temperature centred around clusters and voids. {The Rees--Sciama effect can be estimated from galaxy surveys, and is expected to be important at the largest angular scales \citep{Maturi2007TheUniverse}.}

The slingshot effect is related to the Rees--Sciama effect, but instead of a single hot spot centred on the halo, it creates a dipole pattern with a cold spot in front of and a hot spot behind a moving halo.
\cite{Stebbins2006MeasuringLSS} state that transverse motions of galaxies can be measured in the small scales of the CMB by statistically analysing these dipole patterns. The intention of this paper is to propose a method to measure this effect. {Sufficiently precise measurements of the effect can be used as a self consistency check of our models of structure formation, and as a more direct measure of the gravitational forces acting on large scales.} In the literature, the slingshot effect is also called the Birkinshaw--Gull effect or the moving lens effect. Some times the name of the Rees--Sciama effect is used interchangeably with the slingshot effect, even though it describes a related but slightly different phenomenon.

\begin{figure}
    \centering
    \includegraphics[width=\columnwidth]{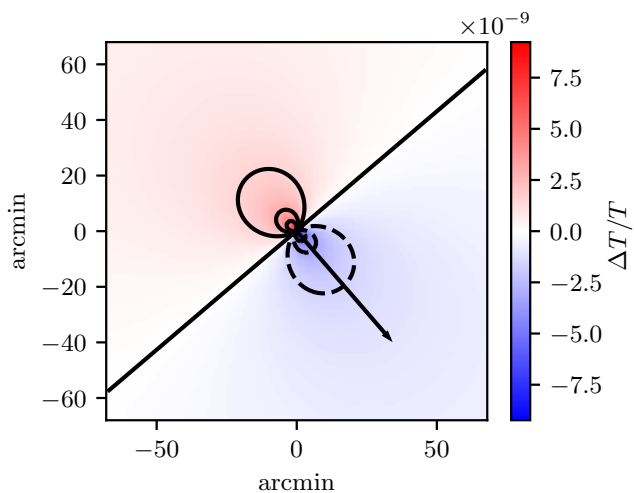} 
    \caption{Contour plot of the temperature dipole created by the slingshot effect. The arrow indicates the direction of motion of the galaxy. An increase in photon temperature is indicated with red colour. A decrease in temperature is illustrated with dashed contours and blue colour. This example is a galaxy of virial mass $1.2\times10^{12}\,M_\odot$, moving with a transverse velocity component of 880 km/s, viewed at a distance of 100 Mpc}
    \label{fig:contour_single}
\end{figure}

 The mechanism behind the dipole pattern of the slingshot effect can be understood as follows: If a CMB photon enters in front of the moving halo, the gravitational potential becomes deeper while the photon is passing through the potential well, meaning that the photon has to spend more energy getting out and becomes redshifted. Likewise, if a CMB photon enters behind the moving halo, the gravitational potential along the photon trajectory becomes shallower while the photon passes, allowing the photon to gain some energy. The result is a dipole signal; An example of the signal is represented visually in figure \ref{fig:contour_single}. The increase or decrease in photon energy can be compared to the gravity assist manoeuvre, where a spacecraft can gain velocity relative to the heliocentric reference frame by passing in the trail of a moving planet. An effect related to the slingshot effect was proposed by \cite{Molnar2013TangentialRedshifts}, where a difference in redshift between two lensed images \emph{from the same source} can be used to infer the tangential velocity of the lens. Other attempts at indirectly inferring transverse velocities include analysing microlensing parallaxes in very specific setups, where one can find a quasar behind the galaxy whose velocity is to be found. See for instance \cite{Gould1994TransverseParallaxes}.
 
The slingshot signal grows stronger with more massive and faster moving structures. \cite{Birkinshaw1983AGalaxies} predicted that a massive and rapidly moving galaxy cluster should produce a measurable signal, but to our knowledge the effect has not been measured yet. As indicated by \cite{Stebbins2006MeasuringLSS}, the difficulty of measuring the effect could be due to similar dipole patterns being produced by the kinematic Sunyaev--Zel'dovich effect and lensing of the CMB anisotropies. With the CMB experiments of today, it is therefore impossible to measure the slingshot effect of a single galaxy. Recently, \cite{Hotinli2018TransverseEffect} investigated the detectability of the slingshot effect by estimating the corrections to the CMB power spectrum. Furthermore, \cite{Yasini2018PairwiseEffect} proposed an estimator for the pairwise peculiar velocities of clusters using the slingshot effect. Both of these recent studies found that the combined effects of transversely moving objects should be measurable with near-future CMB surveys.

The process of image stacking has the ability to isolate the slingshot effect from that of the kinematic Sunyaev--Zel'dovich effect and the lensed CMB. This is because these two confounding effects are not correlated with position and velocity in the same way---averaging their contribution to zero when stacking enough images.
\cite{Maturi2007AstrophysicsGalaxies} propose that one can measure the slingshot effect by stacking the CMB maps of 1000 cluster mergers. In the current work we present a method to detect the average peculiar infall velocity of galaxies around a single cluster. By aligning and stacking the signal from galaxies in a mock catalogue, and the subsequent fitting of the stacked image to a template, we show that the effect is detectable with near future CMB experiments such as CMB-S4 \citep{Abazajian2016CMB-S4Edition}.

\section{Methods}
The ISW effect changes the temperature of the CMB photons during the time $t$ 
they spend in an evolving gravitational potential $\Phi$. Specifically,
\begin{equation}
\frac{\Delta T}{T}=2\int\dot{\Phi}\,\mathrm{d}t,
\end{equation}
where a dot is a time derivative. We adopt units such that $c=1$.
\\
The slingshot effect is the change in photon temperature due to the
transverse motion of an unevolving gravitational well. Following
a flow of photons travelling through a moving halo, this effect can
be expressed as 
\begin{equation}
\frac{\Delta T_{\mathrm{slingshot}}}{T}=2\int\mathbf{v}_{\perp}\cdot\nabla\Phi\,\mathrm{d}t.
\end{equation}
\\
Now we choose a coordinate system where light moves in the positive $z$-direction, along the line of sight. The projected motion $\mathbf{v}_{\perp}$ of the halo is taken to be constant and along the $x$-axis. In this frame, we define $v_x$ as
the velocity component perpendicular to the line of sight.
Since the photons move with the speed of light and we have chosen units
where $c=1$, the time integral can be changed into an integral along
$z$, namely
\begin{equation} \label{slingshot}
\frac{\Delta T_{\mathrm{slingshot}}}{T}=2v_{x}\int\frac{\partial\Phi}{\partial x}\,\mathrm{d}z.
\end{equation}
The coordinates $x$, $y$ and $z$ represent physical distances (measured in non-comoving megaparsecs).
To estimate the magnitude of the slingshot effect, we assume a {simple, yet realistic,} model for the gravitational potential $\Phi$ derived from the halo model setup.

\subsection{Halo model setup}

For the purpose of calculating the signal from a single halo, we model
the halo as a spherically symmetric matter distribution, centred at $x=y=z=0$.
We assume that all of the dark matter halo mass is in a {Navarro--Frenk--White (NFW)} profile with concentration $c_\mathrm{NFW} = 15$ \citep{Navarro1995TheHalos}.
We populate the halo with baryons, consisting of an additional 10\%
of the dark matter mass. We use a Hernquist profile for the
baryons \citep{Hernquist1990TheBULGES}, applying a Hernquist scale length $a$ that relates to the virial radius $r_\mathrm{vir}$ of the dark matter halo according to
\begin{equation}
a=\frac{0.015\,r_\mathrm{vir}}{1+\sqrt{2}}.  
\end{equation}

See the appendix for the detailed calculations of the slingshot effect from the NFW
and the Hernquist component. The resulting expression we implement
to calculate the slingshot effect from a single halo, is
\begin{equation} \label{haloslingshot}
\frac{\Delta T_{\mathrm{slingshot}}}{T}=\frac{2 Gm_{DM} v_{x}}{r_\mathrm{vir}}\left(Q_{\mathrm{NFW}}+\frac{1}{10}Q_{\mathrm{Hernq}}\right),
\end{equation}
where
\begin{equation}
Q_{\mathrm{NFW}}\equiv\frac{gx_r}{x_r^{2}+y_r^{2}}
\left[
\ln\left(\frac{c_\mathrm{NFW}^{2}\left(x_r^{2}+y_r^{2}\right)}{4}\right)
-S\left(2\arctan\left(S\right) - \pi\right)
\right],
\end{equation}
and
\begin{equation}
Q_{\mathrm{Hernq}}\equiv\left( \frac{1 + \sqrt{2}}{0.015} \right) \frac{x_a}{x_a^2 + y_a^2 - 1}
\left[
2
+U\left(
2\arctan\left(U\right)
-\pi
\right)
\right].
\end{equation}
Here, $m_{DM}$ is the virial mass of the dark matter halo. The constant $g$ depends on the concentration $c_\mathrm{NFW}$ according to equation \eqref{eq:g} in the appendix.
We use the following notation for dimensionless coordinates:  $x_r \equiv x/r_\mathrm{vir}$ and $x_a \equiv x/a$. Furthermore, we have defined
 \begin{equation}
     S \equiv \frac{1}{\sqrt{c^{2}\left(x_r^{2}+y_r^{2}\right)-1}},
 \end{equation}
and
\begin{equation}
    U \equiv \frac{1}{ \sqrt{x_a^{2}+y_a^{2}-1}}.
\end{equation}

We emphasise that the $Q$-expressions here are independent of halo parameters for the chosen model. {This means that we only need to} calculate a template of the slingshot effect once, using units of the virial radius, then re-scale the template to galaxies of any size. The deciding factor for the amplitude of the slingshot effect is the combination $m v_x/ r_\mathrm{vir}$.

\subsection{Realistic observational setup}

The slingshot effect, described by equation \eqref{haloslingshot}, increases with the mass of the halo, and with the transverse velocity $v_x$ relative to the CMB.
We find that the slingshot signal from {a single large galaxy} with mass $10^{13} M_\odot$, moving at 1000 km/s,
is around $\Delta T/T\sim10^{-8}$, equivalent to $0.03\,\mu\mathrm{K}$. Most galaxies would be less massive and move slower, {giving an average signal of $\Delta T/T\approx 3 \times 10^{-9}$ (as we will see in section \ref{sub:stack})}. Figure \ref{fig:contour_single} shows the raw slingshot signal from an example galaxy, with a cold spot in front and a hot spot behind the moving galaxy. This example galaxy is slightly heavier and moves slightly faster than the average, resulting in a stronger signal than $\Delta T/T\approx 3 \times 10^{-9}$.

The Atacama Cosmology Telescope (ACT) has instrument noise down to $6\,\mu\mathrm{K\cdot arcmin}$, or a per-pixel noise of $\sigma_{\Delta T/T}\sim 4 \times 10^{-6}$ when assuming 0.5 arcmin pixels\footnote{It is common to construct maps using 2-3 pixels per FWHM of the instrument beam. The FWHM of ACT is 1.3 arcmin.}. See \cite{Hincks2010THEMAPS} for some details on the expected sensitivity and beam profiles {point spread function} of ACT.
The given noise level indicates that {measuring the signal from a single, massive galaxy} is impossible, since the signal of $\Delta T/T\sim10^{-8}$ would drown out in noise. To get a Signal-to-Noise Ratio (SNR) of order one, 
we need to increase the signal---or reduce the noise---by a factor
of 400. The signal increases with the square root of the number of stacked
images, which means that we must stack the signal of over 160 000 fast and massive
galaxies, or closer to $10^7$ average galaxies. Furthermore, these galaxies must be close enough to us for the dipole to be resolved in the CMB map.
The ACT has 1.3 arcmin FWHM beam resolution.  A typical galaxy with virial radius of 100 kpc must be within about 250 Mpc to be resolved with two or more pixels.

An important consideration when stacking these dipole images is that they need
to be oriented correctly, such that the galaxy velocity directions point
along approximately the same axis. Because there are no good ways to measure 
the transverse velocity of a single galaxy yet, one should only stack observable galaxies where the approximate direction of motion can be inferred from other means. While other papers (like \cite{Yasini2018PairwiseEffect}) suggest using the relative velocity of pairs of galaxy clusters, we are in this work studying the infall velocity of galaxies in the large scale structure of the cosmos.
To estimate a direction of peculiar motion, we assume that galaxies will fall towards not yet virialised cluster structures in their vicinity. This is of course not true for absolutely all galaxies, but when stacking images it is sufficient for the sample to have an average velocity in the radial direction. Another option not covered in this paper, is to stack galaxies near the edges of voids; We expect void galaxies to have an average velocity out of the underdense region.

{ We suggest to use the Coma Cluster for aligning the velocities of nearby galaxies in a realistic observational setup. At a distance of about 100 Mpc \citep{Liu2002InfraredConstant}, the Coma Cluster is within the 250 Mpc needed to resolve a slingshot dipole. It is well studied, and not obscured by the Milky Way disk. The Coma Cluster is located approximately at RA of 195 degrees and DEC of +28 degrees. Several current surveys cover this region, for instance the BOSS spectrographic survey of the SDSS (Sloan Digital Sky Survey). 
The Coma Cluster has a mass of about $1.9 \times 10^{15} M_\odot$ (found through weak lensing by \cite{Kubo2007TheSurvey}). Together with the Leo Cluster, it is part of the rich Coma Supercluster, which is large enough to  have not yet completely virialised.
The total amount of galaxies falling towards the supercluster could be sufficient for achieving a SNR of order unity through stacking} \footnote{Assuming that there is on average a few galaxies per cubic megaparsec of space, there should be of the order of $10^6$ galaxies within a 50 Mpc sphere.}. {Reaching an SNR of one requires that all of the infalling galaxies have been catalogued with position and redshift, and that they are on average falling in fast enough.} Furthermore, the presence of foreground sources in our own galaxy can interfere with the precise measurement of the CMB around some of these galaxies.

To identify the signal even when the total stacked SNR is less than one, we propose to fit the
expected dipole signal to the stacked image. {We will furthermore estimate the probability of having false positives by fitting a template of the expected signal to several stacks of uncorrelated noise maps.}
If the best fit dipole is sufficiently improbable to achieve randomly with
just noise, we can say that the dipole is detectable, even if it is not visible by eye in the stacked image.

\subsection{Stacking images from a simulated halo catalogue}\label{sub:stack}

To emulate realistic halos, we use the halo catalogue from the MultiDark Planck 2 survey, described by \cite{Prada2012HaloCosmology}.
The data set contains dark matter halos identified with Rockstar halo finder \citep{Behroozi2013TheCores}. 
In this $\left (1 \mathrm{Gpc}/h \right )^3$-box simulation, we
identify one of the largest halos. It has virial mass in the order
of $M_\mathrm{vir,\,supercluster} \approx 10^{15}\,M_{\odot}$ and virial radius $r_\mathrm{vir,\,supercluster} \approx$ 4 Mpc, {which is similar to the Coma Cluster}. We find that this is the centre of a
massive supercluster that has not yet virialized, meaning that nearby
smaller halos are falling towards the supercluster. The virial mass and virial radius estimated by the halo finder at redshift zero is significantly lower than the actual mass affected gravitationally.

\begin{figure}[t!]
    \centering
    \includegraphics[width=\columnwidth]{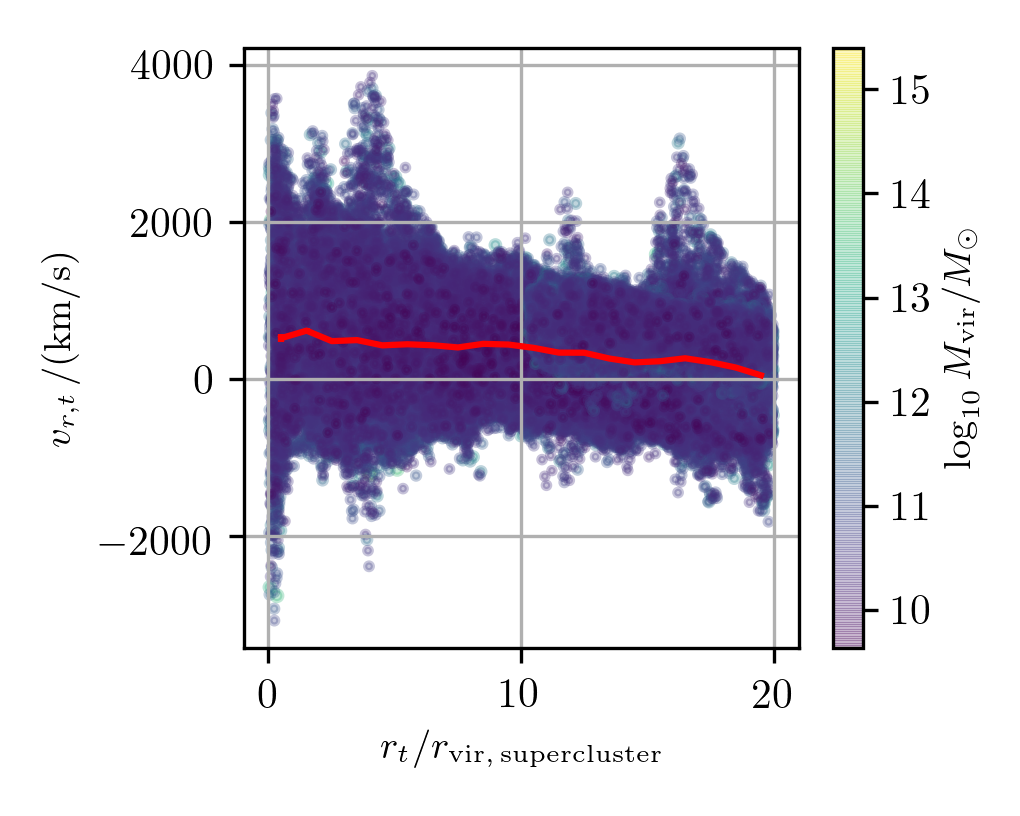}
    \caption{Scatter plot of the projected radial motion of halos close to the massive supercluster. The horizontal axis shows the projected distance from the centre of the structure (in units of the virial radius of the supercluster). The vertical axis shows the projected radial velocity of the halo in question, with positive values indicating infalling motion. The red line shows the binned mean. {The error of the mean is consistent with the width of the line.} The colour of a point indicates the virial mass of that halo.}
    \label{fig:velocity_scatter}
\end{figure}

When studying subhalos close to the identified supercluster, we find
that they indeed have a velocity component directed towards the centre of
the structure. See figure \ref{fig:velocity_scatter} for a scatter plot of the velocities of halos surrounding the supercluster. {The net inward flow of matter is apparent from closer than $1\, r_\mathrm{vir,\,supercluster}$, up to distances of about $20\,r_{\mathrm{vir,\,supercluster}}$. }

We pick out nearby halos with virial mass above $10^{11}M_{\odot}$
within 10 $r_{\mathrm{vir,\,supercluster}}$ (approximately 40 Mpc) of the central supercluster. The mass cutoff is chosen to exclude halos hosting dwarf galaxies, which are more difficult to observe and add a weak signal to the stack. The distance cutoff was chosen by hand; Repeating the analysis with a 15 $r_{\mathrm{vir,\,supercluster}}$ distance cutoff instead does not significantly improve the results compared to 10 $r_{\mathrm{vir,\,supercluster}}$. We draw the conclusion that this is because the halos further away add more noise than signal, for instance because their velocities are misaligned due to other nearby structure. 
The chosen cut of 10 $r_{\mathrm{vir,\,supercluster}}$ and $M > 10^{11}M_{\odot}$ leaves 8438 halos in our final sample. These halos mostly
consists of galaxies, but also some galaxy groups and galaxy clusters.
The selected halos have an average mass of $1.79\times10^{12}\,M_\odot$. 
We choose the $z$-axis of the 3D simulation as the line-of-sight,
and project the velocity of each halo down to the plane perpendicular
to this axis (the $x$-$y$ plane).

To simulate the signal observed with ACT, we create a 2D map of $\Delta T_{\mathrm{slingshot}}/T$ for each galaxy halo. The map size is chosen to fit two times the virial radius of that galaxy, with 0.5 arcmin pixel size.
We smooth the map with Gaussian blur with 2.6 pixels
FWHM to emulate the 1.3 arcmin beam of ACT \citep{Hincks2010THEMAPS}.
We assume that each pixel of each image has an independent noise, 
drawn from a normal distribution with standard deviation $\sigma_{\Delta T/T}$. 
We perform the full analysis with three different values for the noise standard deviation: 
$4\times10^{-6}$, $1\times10^{-6}$, and $1\times10^{-7}$. 
We do not vary the other parameters, such as the angular resolution.

{We neglect perturbations in the CMB background, because we assume the background can be well enough modelled and subtracted on the relevant scales (corresponding to $l > 500$).
We also neglect non-dipole contributions like foreground sources. All of these are either expected to average out when stacked (if they are not correlated with the galaxy position and direction of motion), or have a monopole signal which will not contribute to the dipole stencil fit. Furthermore, we also neglect galaxy lensing of large scale CMB gradients, which will introduce dipoles on similar scales as the slingshot effect. The large scale gradients of the CMB are not expected to be correlated with the velocity of low redshift galaxies falling into a massive cluster, so the amplitude of lensing of the CMB will be negligible with large enough data sets, as outlined by \cite{Stebbins2006MeasuringLSS}. To further help remove this confounding signal, one can apply a delensing algorithm to the map around each galaxy before stacking (see e.g. \citealt{Manzotti2018FutureSurveys}. A process of delensing for the purpose of isolating the slingshot signal is suggested in \citealt{Maturi2007AstrophysicsGalaxies}). Similarly, we also neglect the signals from nearby galaxies or galaxies located behind the observed ones; if another galaxy is within $2 r_\mathrm{vir}$ of the imaged galaxy, it will introduce an additional dipole which is not positioned at the centre of the map. The chance of such an overlap is not negligible, but since the relative positions are not considered correlated with the direction of infall velocity, the stacking and fitting process is not expected to be sensitive to this signal.}

\begin{figure*}[ht]
    \centering
    \includegraphics[width=0.33\textwidth,trim={1cm 0 0.5cm 0},clip]{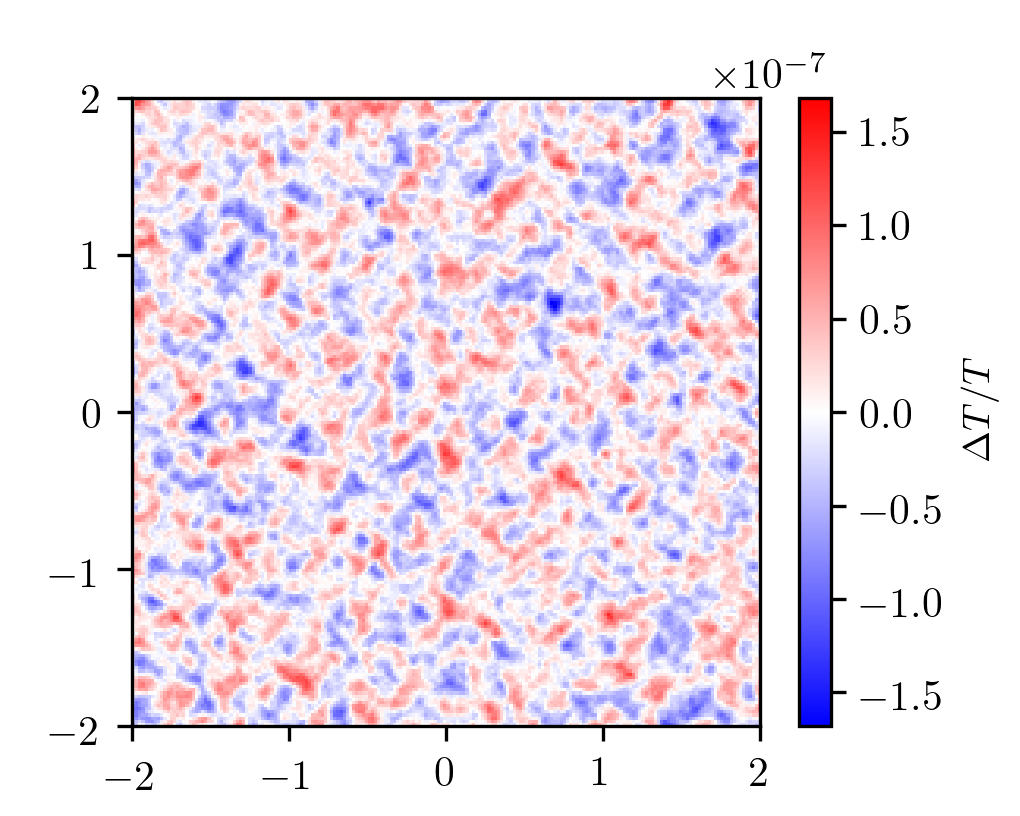}
    \includegraphics[width=0.33\textwidth,trim={1cm 0 0.5cm 0},clip]{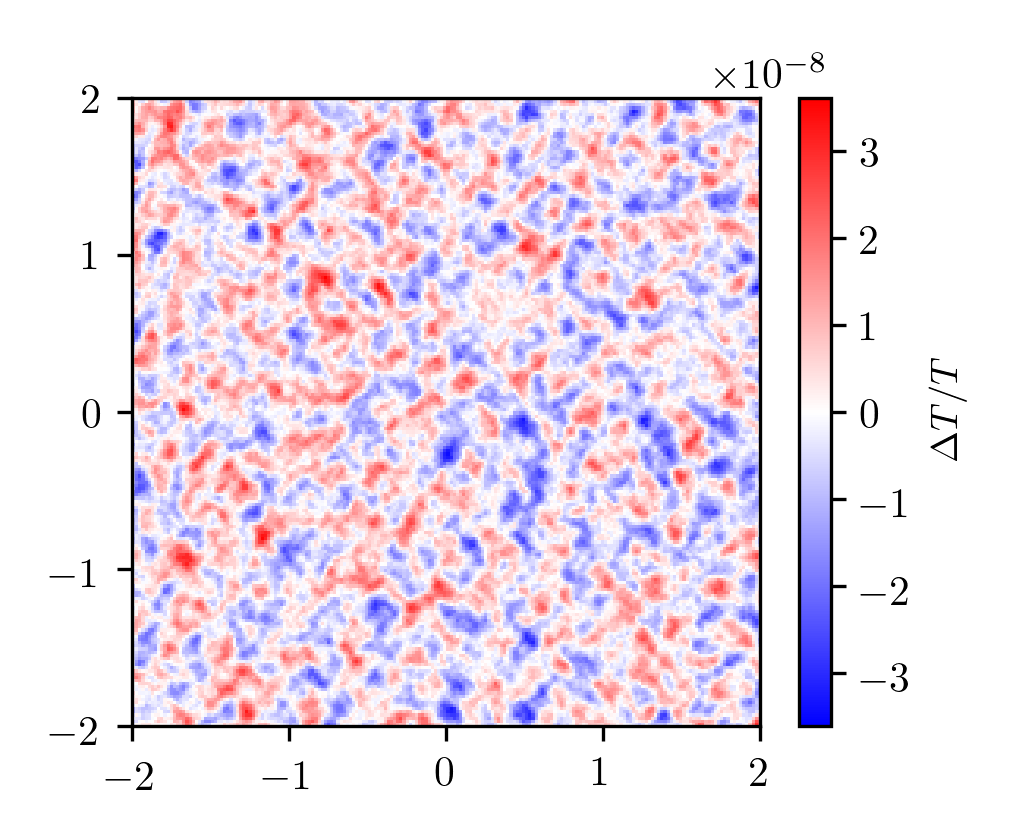}
    \includegraphics[width=0.33\textwidth,trim={1cm 0 0.5cm 0},clip]{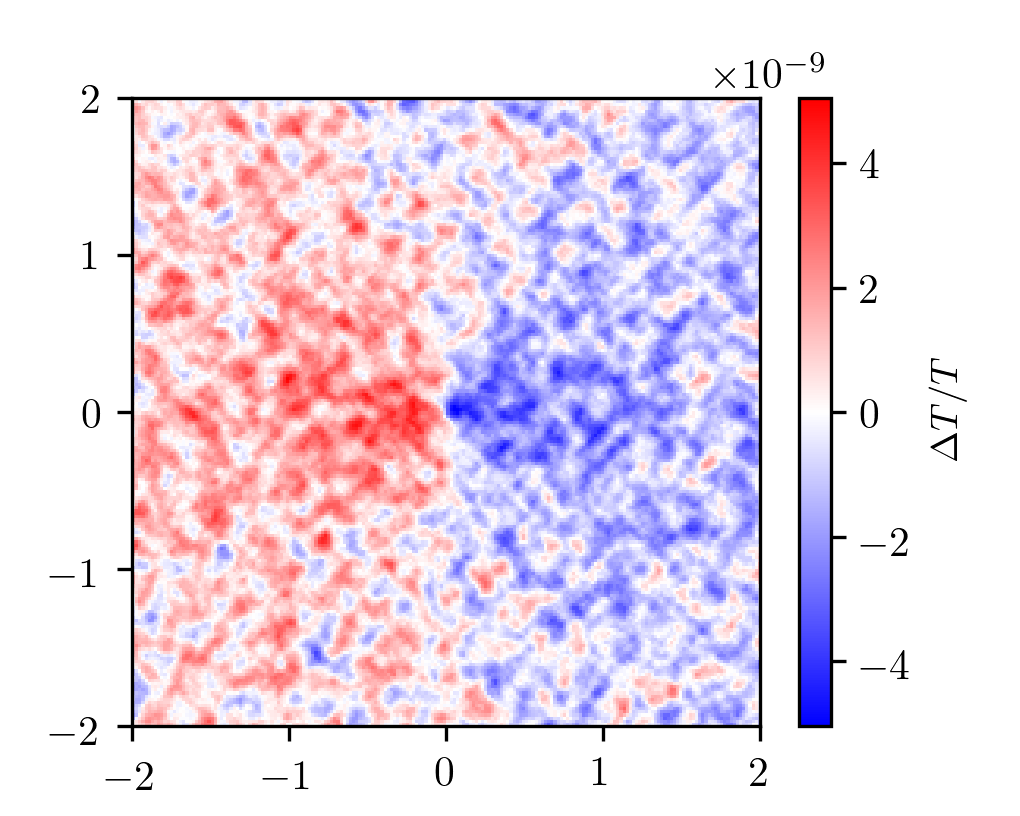} \\
    \includegraphics[width=0.33\textwidth,trim={1cm 0 0.5cm 0},clip]{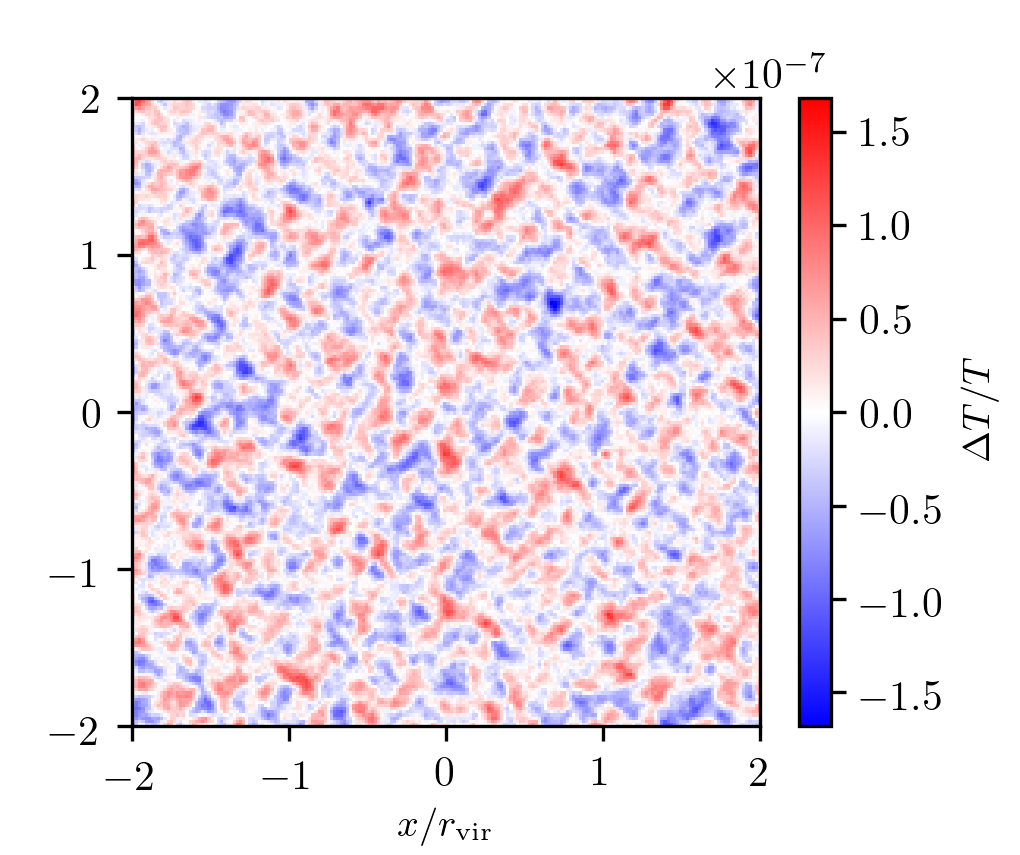}
    \includegraphics[width=0.33\textwidth,trim={1cm 0 0.5cm 0},clip]{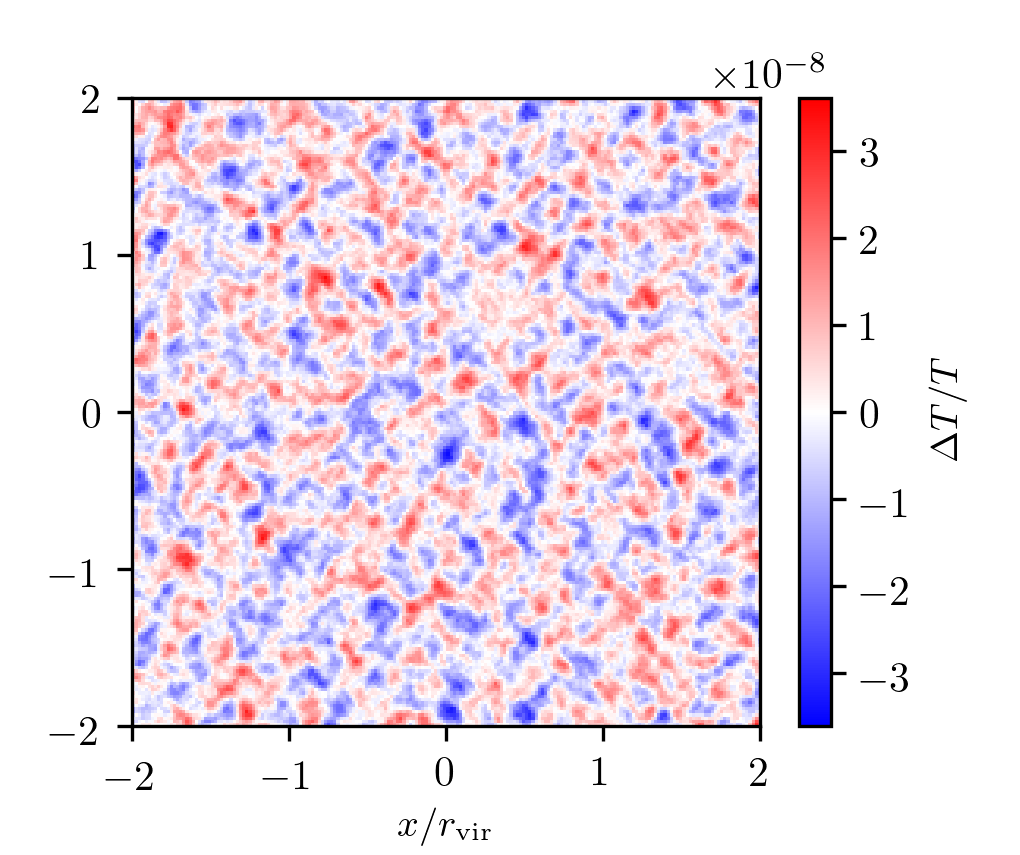}
    \includegraphics[width=0.33\textwidth,trim={1cm 0 0.5cm 0},clip]{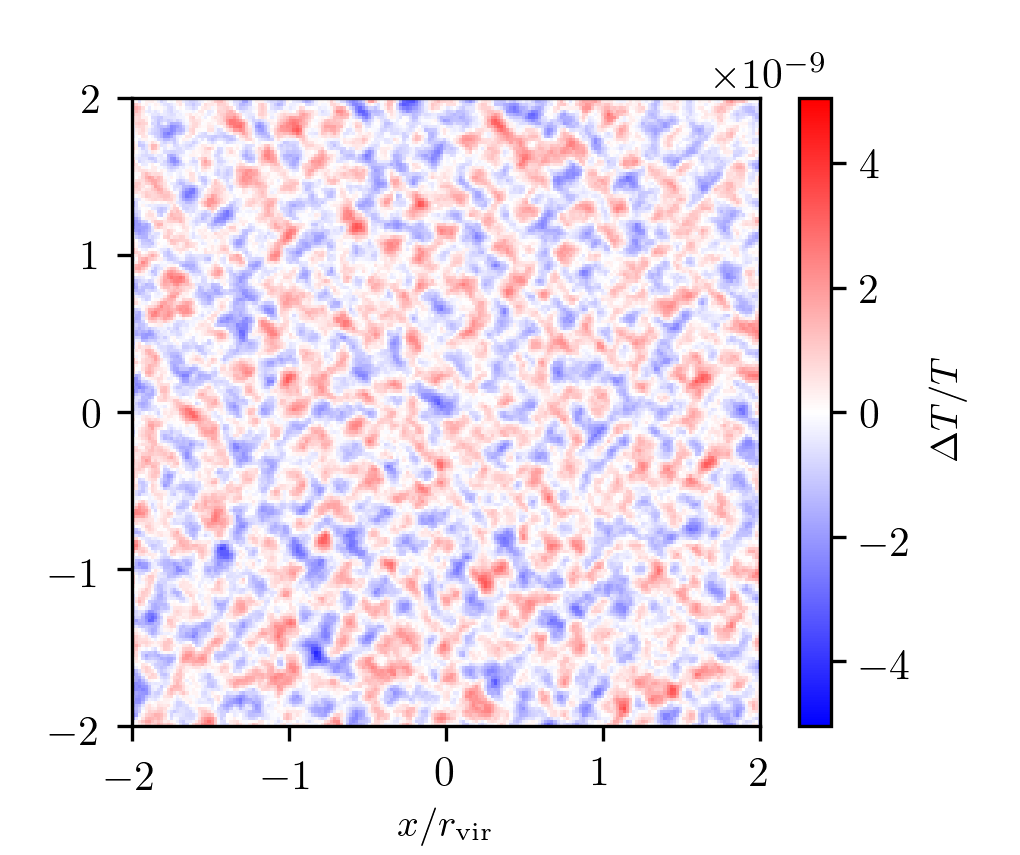} \\

    \caption{Example realisations of the stacks (top row), and the residuals (bottom row) after stacking the slingshot signal of 8438 halos {and subtracting the best fit. }
    The left panels are for a noise per pixel of $\sigma_{\Delta T/T} = 4\times10^{-6}$, resembling current surveys. The central panels have near-future noise levels of $\sigma_{\Delta T/T} = 1\times10^{-6}$, and the rightmost panels are for a very low noise per pixel ($\sigma_{\Delta T/T} = 1\times10^{-7}$) to visually see the emergence of an image. The individual images are scaled with respect to the virial radius of the halo before stacking. The $x$- and $y$-axes are in units of the virial radius of the halo. }
    \label{fig:stack_fit}
\end{figure*}

{The combined maps of each nearby halo are, after {applying the instrumental point spread function} and adding noise, aligned according to the expected infall direction, radially towards the centre of the supercluster. This is to simulate real observations, where we have no a priori knowledge of the individual peculiar velocities. Since we use realistic simulations, our mock data sets take into account the fact that the individual galaxy motion is not necessarily aligned with the radial direction.} Before stacking, the maps are re-scaled with respect to the virial radius of each galaxy, using linear interpolation. This ensures that the stacked image will not be radially smeared due to size inconsistencies. In the main analysis we assume that we have perfect knowledge of the virial radius, but we also do a smaller analysis {consisting of fewer individual stacks than the main analysis, where we include the uncertainties on} the virial mass and radius. This is to show that our results are robust to this uncertainty. 

The signal from each individual halo is summed up, and divided by the number of halos to achieve an average (or effective) signal. Example realisations of this average signal after stacking 8438 halos are showed in the top row of figure \ref{fig:stack_fit}.
The three columns are for three different noise levels: current ACT noise ($\sigma_{\Delta T/T} = 4\times10^{-6}$), a near-future CMB-S4 noise level ($\sigma_{\Delta T/T} = 1\times10^{-6}$), and a futuristic survey ($\sigma_{\Delta T/T} = 1\times10^{-7}$). We do not vary the resolution between these experiments, because CMB-S4 is expected to have a similar spatial resolution to ACT (1 and a few arcminutes, see \citealt{Abazajian2016CMB-S4Edition}). {The ACT noise level is equivalent to $6\,\mu\mathrm{K\cdot arcmin}$ \citep{Hincks2010THEMAPS}, while the CMB-S4 noise level is equivalent to $1.5\,\mu\mathrm{K\cdot arcmin}$, consistent with the $\sim 1\,\mu\mathrm{K\cdot arcmin}$ \citep{Abazajian2016CMB-S4Edition}. }
The futuristic noise level in the rightmost column is shown to emphasise the expected visual signal if we had sufficiently good statistics, and is not considered realistic in the foreseeable future. {The best fit stencil to the low noise stack suggests a signal amplitude of $\Delta T/T \approx 3 \times 10^{-9}$.}
We are stacking almost 10 000 halos, which is expected to increase the Signal-to-Noise by a factor of 100. The best fit and the residuals are of similar amplitude when the noise per pixel is $10^{-7}$, indicating a SNR $\sim 1$ after stacking. The stacked signal is expected to be 100 times stronger than the average signal from a single halo, which confirms that an average halo has a signal of the order of $\Delta T/T \sim 10^{-9}$. {The realisation with near-future noise level, shown in the middle column of figure \ref{fig:stack_fit}, also recovers the expected signal with amplitude $\Delta T/T \approx 3 \times 10^{-9}$, but the ACT noise level of $4 \times 10^{-6}$ results in a too low SNR for the fitting algorithm to find a statistically significant dipole pattern in the shown example.}

{In the main analysis we use a uniform weighting when averaging the stacks. However, the SNR can possibly be improved by choosing a different weighting for the galaxy maps. Specifically the weighting can be chosen in such a way that bigger and faster galaxies contribute more to the stack than small and slow galaxies, which add mostly noise. As seen in figure \ref{fig:velocity_scatter}, the average radial velocity is expected to be greatest around 1 virial radius from the central region of the supercluster, then reduce with distance. This suggests the possibility of weighting the galaxies according to distance. Another possibility is to weight them according to their estimated mass.
In addition to the main analysis, we perform a small analysis with three weighting schemes: one weights galaxies according to projected distance from the centre of the supercluster, $r_{\perp}$ (normalised to the virial radius of the supercluster), with weight $w_d = \sqrt{r_{\perp}} e^{-0.3 r_{\perp}}$. The other weights galaxies according to the logarithm of the galaxy mass (including a random spread in the halo mass) as $w_m = \log_{10}(M_\mathrm{vir}/ 10^{10} M_\odot)$. The last weighting scheme combines the distance and mass weighting, $w_c = w_d \times w_m$.}

\subsection{Observational Challenges}
{The analysis assumes we can identify $\mathcal{O}(10^4)$ galaxies near the Coma cluster. The area of the sky where these galaxies can be found is a 40 Mpc radius circle at a distance of 100 Mpc, corresponding to 1400-1500 square degrees.
There is a possibility that the least massive galaxies will not be visible in the galaxy surveys. As a rough estimate for the observed magnitude of the faintest galaxies studied in this paper, we assume $10^{11} M_\odot$ halo mass, 1/10 mass in baryons, and a mass-to-light ratio of 10 {(in \cite{Faber1979MassesGalaxies}, all of the measured galaxies are found to have lower mass-to-light ratio than 12)}. This results in a luminosity of $10^9$, and a bolometric magnitude of approximately 19 at the largest distances of up to 200 Mpc. This is within the scope of contemporary surveys, like the SDSS. 
The slingshot signal contribution will be strongest from the most massive galaxies, so missing a few of the faintest galaxies is not expected to change the stack significantly.}

{To rescale the galaxy maps correctly, we suggest to infer the virial radius from the galaxy virial mass. In many cases---when a lensing analysis has not been performed---the total halo mass is not known, and must be estimated from the galaxy luminosity. A possibility is to go via the mass of neutral hydrogen (HI), which can be measured from the HI line, or estimated from visible light.
To distinguish galaxies in the vicinity of the cluster from distant background galaxies, it is important to know the redshift of each galaxy fairly well, within 50\% error or so. At the distance of 100 Mpc (corresponding to about $z=0.02$), the uncertainty in photometric redshift is $\sigma_z \gtrsim 0.03$ \citep{Bolzonella2000PhotometricProcedures}. This implies that we need spectroscopic data for all the galaxies used in the stacking process.}

{Several of the above problems can be partially solved if we consider data from the upcoming Square Kilometre Array (SKA) phase 2, expected to be online in 2030 \citep{Bull2016EXTENDINGARRAY}.
SKA is a planned full-sky spectroscopic survey, and \cite{Norris2014TheRadio-astronomy} suggest that it is expected to identify the position and redshift of all galaxies of the relevant magnitudes up to $z=0.05$. We expect SKA to find the HI mass of all viable slingshot galaxies with fair certainty. }

{Whether using data from SKA or the SDSS, the biggest source of error when estimating the virial radius is the conversion from HI mass to halo mass. \cite{Villaescusa-Navarro2018IngredientsMapping} and \cite{Padmanabhan2017ConstraintsGyr} both indicate about one order of magnitude spread in the halo mass for a given HI mass. We do not include this error when rescaling the individual maps in the main analysis, but we perform a separate, smaller analysis where we include the corresponding error in virial radius. We find that the error in rescaling of the virial radius does not induce a bias, and does not significantly increase the uncertainty of the results. We expand upon this in the results section (section \ref{sec:results}).}

In this analysis, we assume that the CMB behind the galaxies
is known and can be subtracted. If there are no other significant contributions between
the surface of last scattering and the observed galaxy, the CMB is a Gaussian
random field with a standard deviation in $\Delta T/T$ of $\sigma_{\Delta T/T}=10^{-4}$.
By masking out the signal surrounding the galaxy, one can analyse the CMB signal from the external area, and interpolate
it into the masked region. See for instance \cite{Bucher2012FillingRealizations} for an example prescription to fill in masked regions of the CMB map. Subtracting the expected interpolated CMB from the actual
observed signal will leave signal and noise, {without the CMB perturbations.
Likewise, the galaxy itself, lensing of gradients in the CMB, and foreground sources can outshine the dipole signal from the slingshot effect. None of these effects are expected to give a dipole correlated with the infall velocity, so the process of stacking in itself should {suppress} any apparent signals from other sources.}

\section{Statistical analysis}

For realistic noise levels of $10^{-6}$ and above, it is impossible to see the dipole fit by eye.
We compare the stacked image to the model in equation \eqref{haloslingshot}, using a least squares method with two free parameters. {The two free parameters used in the fit are:} the combination $m v_x/ r_\mathrm{vir}$, which gives the amplitude of the signal; and an image smoothing radius $r_\mathrm{smooth}$, which is related to the instrument beam. {When stacking images that are re-scaled with respect to their radius, the stack will consist of images with different effective beam widths. The resulting merged image is not exactly equivalent to a dipole signal with a single Gaussian smoothing like the one we apply in our least squares procedure. We find that fitting the stencil with a single effective smoothing radius consistently over-estimates the signal with 3--4~\%. This can be avoided by choosing a different stencil, for instance pre-generating a stencil from a stack of noiseless smoothed maps.}

{In the following sections, we introduce measurements of the quality of the fit and amplitude of the signal.  We also show that these two quantities can be combined into a single estimator that can be used to distinguish a true detection from a false positive.}

\subsection{Lower bound on the dipole amplitude}
If the best fit of a stack corresponds to a very low amplitude $m v_x/ r_\mathrm{vir}$, it is indistinguishable from zero amplitude. A detection limit for this number should be estimated based on the error bars of the data, as well as the expected masses and velocities of the stacked galaxies. This limit will therefore depend on the mass and structure of the central supercluster and on the precision of data in the galaxy catalogue in the real-world scenario. {The least squares method we apply has discrete values for the combination $m v_x/ r_\mathrm{vir}$. This means that below some threshold value, the amplitude will be rounded down to zero in our implementation. We choose this value conservatively in a way that does not significantly impact the results.}

In our case, the average expected velocity is $\mathcal{O}(100)$ km/s (as seen in figure \ref{fig:velocity_scatter}), and the average galaxy mass is $1.8\times10^{12}\,M_\odot$. To be conservative, we set the threshold for a zero amplitude detection corresponding to a $10^{12}\, M_\odot$ galaxy, moving at less than 10 km/s. This choice is arbitrary and can be chosen differently when handling actual observations. The chosen threshold is equivalent to a factor of about 0.05 of the expected value for $m v_x/ r_\mathrm{vir}$. This does not mean that we ignore individual images of galaxies that are smaller or slower than this threshold; We stack all galaxies, and consider the stacked signal to have zero amplitude if the best fit indicates that the \emph{average} value for $m v_x/ r_\mathrm{vir}$ is less than $\sim 5\,\%$ of the expected average.

\subsection{Quality of fit}
For each stack $d$, we find the best fit template $t$ with a least square method. The template $t$ is a smoothed image of the pure slingshot signal (equation \ref{haloslingshot}). Both $d$ and $t$ are column vectors containing each pixel of the stacked image and the best fit template.
To gauge the quality of the fit, we calculate the normal equations 
\begin{equation}
    q=\frac{t^T d}{t^T t},
\end{equation}
{where $t^T$ is a row vector equivalent to the transpose of $t$.}
This statistic is related to the $\chi^2$-statistic. If you imagine the $t$ and $d$ vectors of dimension $n = n_x \times n_y$ {(with $n_x$ and $n_y$ being the amount of pixels in the $x$- and $y$-direction of the map respectively)}, the $q$ statistic is equivalent to the Euclidean dot product between the data and the template, normalised to the length of the expected template vector. The result is equal to 1 if the two vectors are of equal length and parallel to each other {and 0 if they are orthogonal}.
Calculating $q$ is equivalent to summing up the pixel-by-pixel product of stack and template (vector dot product), and normalising to the squared norm of the template. 

\begin{figure*}[ht!]
    \centering
    \includegraphics[width=0.33\textwidth]{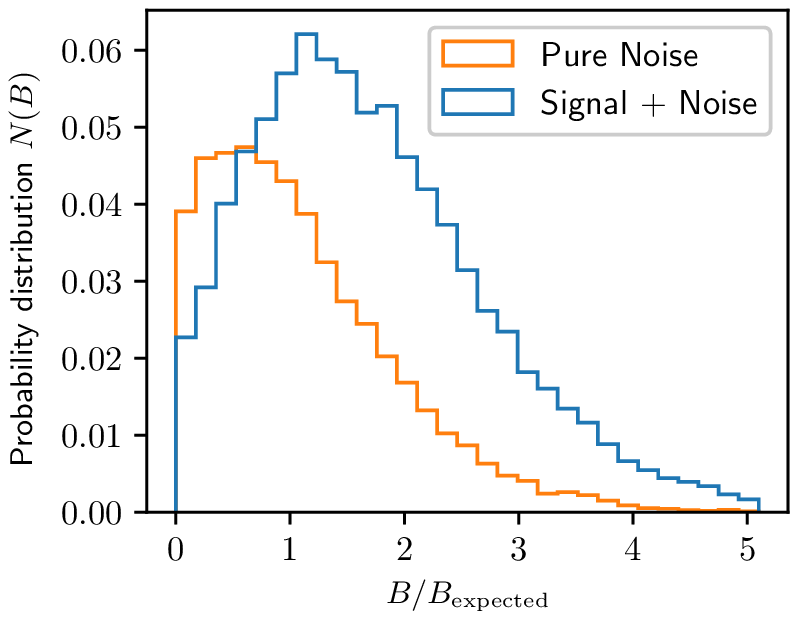}
    \includegraphics[width=0.33\textwidth]{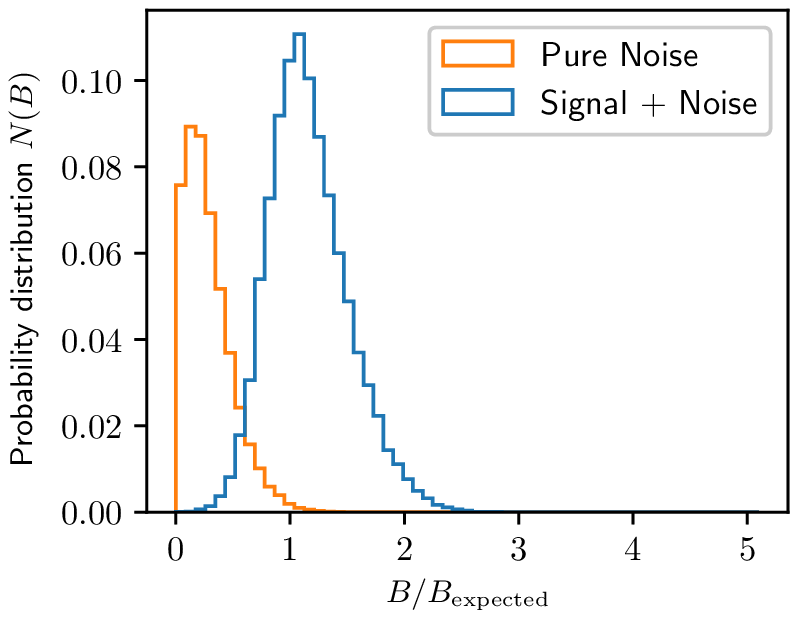}
    \includegraphics[width=0.33\textwidth]{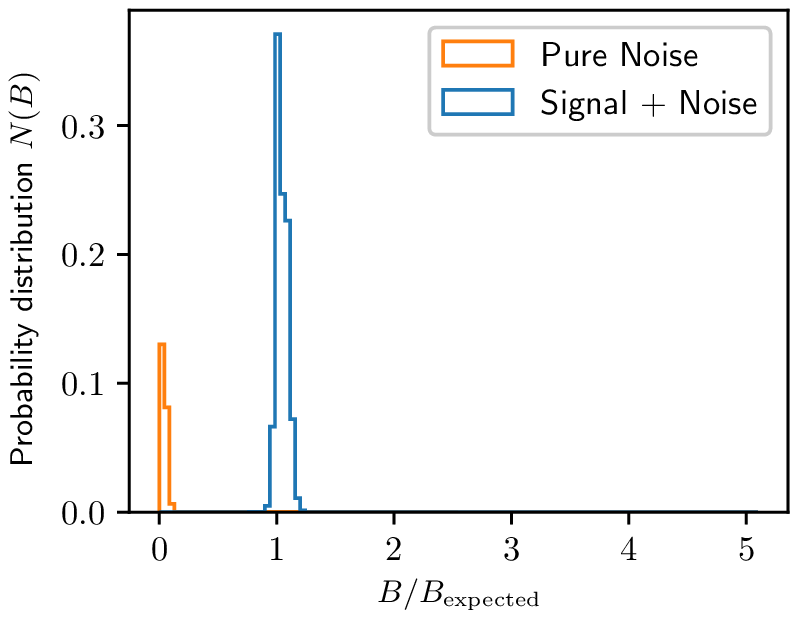}

    \caption{Histogram of the $B$ statistic from stacks of pure noise (orange) and of signal + noise (blue). The three panels correspond to  $4\times10^{-6}$, $1\times10^{-6}$, and $1\times10^{-7}$ noise levels respectively. Values of $B$ below 0 are not plotted.}
    \label{fig:4e6_statistic}
\end{figure*}

A true fit of a stack with a slingshot signal, like the ones shown in figure \ref{fig:stack_fit}, gives $q \approx 1$ if the noise is low. Fitting a template when no signal is present gives $q \approx 0$. The probability distribution of the best fit $q$ has a spread, which widens with higher noise levels. In practice it is possible to get $q > 1$ if the stack data $d$ is parallel and has a bigger amplitude than the template $t$. It is also possible to get $q < 0$ if the dot product is negative (i.e. the vectors are anti-parallel). Both of these options are worse quality fits than $q = 1$.

\subsection{Combined statistic to identify false positives}
In addition to the possibility of a low quality fit when the signal is present, there is also a chance for the algorithm to detect a signal in pure noise with no actual signal present. We will call such a detection a false positive. 
We consider a signal to be a is a proper detection if two conditions are met simultaneously: the quality of fit $q$ is close to 1, and the fitted amplitude of the combination $m v_x/ r_\mathrm{vir}$ is close to the expected amplitude. We consider a signal to be a bad detection if the best fit amplitude is low or if the quality of fit  $q$ is small (or negative). To take into account both of these measures of the detection level, we define a combined statistic,
\begin{equation}
    B \equiv q \times \frac {m v_x}{ r_\mathrm{vir}}.
\end{equation}

The $B$ statistic combines both qualities we are interested in when looking for a good detection: a significant amplitude for the fitted dipole, and a decent quality of the fit. We make a probability distribution of the outcome for measurements of $B$ given a specific noise level, by creating several real stacks and several false stacks (with no dipole signal). 
We create several realisations of the real stack, by repeating the pipeline described in section \ref{sub:stack}, but using a different random seed for the noise for each realisation.
We also create several false stacks, where each of the stacks consists of images of the 8438 halos in our selection. The individual images pass through the same pipeline as the real stacks, but without the addition of the slingshot dipole; Each image consists purely of the per-pixel noise, and is later smoothed and re-scaled with respect to the expected virial radius of the halo.

For each realisation we calculate $B$, and combine the data from all samples to find a probability distribution for $B$. If there is significant overlap between the probability distributions from the stacks with a signal and without a signal, it is impossible to distinguish if an observed stack contains a signal or not. If the distributions are sufficiently separated along $B$, the probability of a strong observed signal being a false positive is small, hence we can distinguish a true signal from a false positive.

\section{Results}\label{sec:results}

{In this section, we first discuss the probability of a statistically significant detection, with current and future surveys. Then we present results from our smaller sets of test analysis, where we estimate the bias from having an uncertainty in the assumed virial radius of a galaxy, as well as evaluating the effectivity of our weighting schemes. Finally we consider how the results apply to modified gravity.}

\subsection{Detection probability}

{Histograms of $B/B_\mathrm{expected}$ are shown in figure \ref{fig:4e6_statistic}.} The three panels correspond to the distributions found from stacks with $4\times10^{-6}$, $1\times10^{-6}$, and $1\times10^{-7}$ noise respectively. 
{The histograms emerging from multiple realisations of the stack can be interpreted as an estimate of the probability distribution of measuring a certain value of $B$, given a known noise level.}
The orange histograms are for stacks of pure noise, while the blue histograms are for stacks of signal and noise. The values are binned in evenly spaced bins from $B = 0$ to $B/B_\mathrm{expected} = 5$. 
{The expected value, $B_\mathrm{expected}$, is found by assuming $q=1$ and taking the average ${m v_x}/{ r_\mathrm{vir}}$ in the sample. The average mass of the halos in our sample is $1.79\times 10^{12}\,M_\odot$, the average radial infall velocity is 675~km/s, and the average virial radius is 196~kpc. For the galaxies around the simulated supercluster we study, we find the value }
\begin{equation}\label{eq:Bexp}
    B_\mathrm{expected} = 7.29\times 10^{9} \,M_\odot/\mathrm{Mpc} = 2.08\times 10^{-10} / G.
\end{equation} 

In the realisations with $4\times10^{-6}$ noise, the histograms of the stacks with signal and with pure noise overlap significantly, and the probability to observe a value we can distinguish from noise is low.
For near-future noise levels of $1\times10^{-6}$, the situation is better. The mean value for $B$ when no signal is present is about three standard deviations lower than the mean value for $B$ with signal. This suggests a very good probability of measuring a value of $B$ big enough to be reasonably sure it is not from noise. 
In the futuristic noise realisations, the histograms do not overlap at all, meaning that in this theoretical scenario we can always distinguish the existence of a signal from a case with no signal.

\begin{table*}[h!]

    \caption{\label{tab:stack_table}Details for the analysis of the simulated stacks} 
    \centering
    \begin{tabular}{cc|ccccc|cc}
        Noise & Signal & $\#$ stacks & $P(\mathrm{ND})$ & mean $q$ & mean $B$ & $\left (m v/r_\mathrm{vir} \right )$ & 95~OSCI $\left( B/B_\mathrm{expected} \right)$ & $P(\mathrm{FP})$\\
         \hline
        $4\times10^{-6}$ & No & 45 000 & 72.7~\% & 0.413 & 0.514 & $0.68 \pm 0.89$  & \, & 51.44~\%\\
        $1\times10^{-6}$ & No & 65 000 & 72.9~\% & 0.399 & 0.130 & $ 0.176 \pm 0.220$ & \, & 2.89~\%\\
        $1\times10^{-7}$ & No & 45 000 & 81.0~\% & 0.1696 & 0.0089 & $ 0.023 \pm 0.018$ & \, & 0.00~\%\\
         \hline
        $4\times10^{-6}$ & Yes & 20 000 & 45.0~\% & 0.898 & 1.433 & $1.59\pm 1.23$ & 0 & \, \\
        $1\times10^{-6}$ & Yes & 20 000 & 0.22~\% & 0.986 & 1.165 & $1.200 \pm 0.374$ & 0.661 & \, \\
        $1\times10^{-7}$ & Yes & 20 000 & 0~\% & 0.9961 & 1.046 & $1.050 \pm 0.052$ & 0.9709 & \,\\
    \end{tabular}
    \tablefoot{$P(\mathrm{ND})$ is the probability of a non-detection (i.e. how many of the stacks are indistinguishable from $B=0$ within $1 \sigma$). The mean measured $B$ and the mean amplitude $m v/r_\mathrm{vir}$ are normalised with respect to the expected value of the average sample.
    The amplitude column shows the average value and also the standard deviation among all the realisations.
    95~OSCI is the lower 95~\% One-Sided Confidence Interval, or the 5th percentile of $B/B_\mathrm{expected}$. The rightmost column shows the probability $P(\mathrm{FP})$ of a false positive among the stacks of pure noise.}
    
\end{table*}

{We present data for the different runs} in table \ref{tab:stack_table}. 
$P(\mathrm{ND})$ is the probability of a non-detection. 
The chance of a non-detection is calculated as the probability for a given stack to be indistinguishable from $B = 0$ within one standard deviation. For noise levels of $4\times10^{-6}$, we find a 45~\% probability for non-detections among the stacks that include a signal. This means that, even with experiments of today, there is a better than 50-50 chance to achieve a fit that is distinguishable from zero. For the near-future noise levels, we get very few non-detections in the stacks that include an actual signal. However, due to the fact that pure noise can lead to a detection as well, we must also consider the chance for a false positive before labelling any detection above the threshold a true detection.

The mean $B$ shown in the table is normalised with respect to the expected value of $B$ for the galaxies in the sample {(i.e. the shown $B$ is divided by $B_\mathrm{expected}$ from equation (\ref{eq:Bexp})). Specifically, unbiased measurements of the normalised $B$ have an expected mean of $B = 1$ when the dipole signal is present, and 0 when there is only noise.}
The amplitude $m v/r_\mathrm{vir}$ of the signal is also normalised with respect to the expected amplitude, $\left (m v/r_\mathrm{vir} \right )_\mathrm{expected}$, {which has the same value as $B_\mathrm{expected}$}.
{The column labelled} 95~OSCI contains the lower 95~\% One-Sided Confidence Interval, or the 5th percentile of $B/B_\mathrm{expected}$. 95~\% of all stacks in this set of stacks has a $B$ above this level. 
The probability of a false positive, $P(\mathrm{FP})$, {is shown in the last column of table} \ref{tab:stack_table}. This probability is found by computing how large percentage of the stacks with pure noise and no signal that give a value of $B$ that falls above the 5th percentile we would expect if there was a signal. Specifically $P(\mathrm{FP}) = P(B_\mathrm{no-signal} > B_{95 \, \mathrm{OSCI}})$ where $B_\mathrm{no-signal}$ is $B$ in a stack with pure noise, and $B_{95 \, \mathrm{OSCI}}$ is the lower 95~\% OSCI for $B$ in the stacks with signal.

A conservative estimate for the chances of measuring a false positive can be found in the rightmost column of table \ref{tab:stack_table}. We focus on the results for the near future experiments with $1\times10^{-6}$ noise. A true stack with this noise level has $B/B_\mathrm{expected} > 0.661$ with 95~\% confidence. If there is no signal in the stacks, the chance of getting $B/B_\mathrm{expected} > 0.661$ is just 2.89~\%. This means that, with CMB-S4, we can distinguish the slingshot signal from pure noise with $P<0.05$ certainty. Furthermore, if the signal is present in near-future observations, we should also be able to put error bars on the measurement of the combination $m v/r_\mathrm{vir}$, because the expected value for this combination is approximately three standard deviations away from zero.

\subsection{Testing assumptions and weighting schemes}
{We repeat four sets of a smaller analysis, where each set includes 5000 realisations of the stack instead of the 20 000 realisations used in the main analysis. One of the sets includes a random error in the assumed virial masses, which induces an error in the virial radius when assuming $m_\mathrm{vir} \propto r_\mathrm{vir}^3$. The result of including this error is that the stack is slightly smeared, but the dipole stencil still detects the signal well. For the $10^{-6}$ noise level simulations, the chance of detecting a false positive increases from 2.89~\% to 4.06~\%, and the standard deviation of the recovered $m v/r_\mathrm{vir}$ increases slightly. The detection is still greater than two sigma significance, indicating that the analysis is robust to the possible error in estimated virial radius.}

{The other three sets of analyses include the three non-uniform weighting schemes described in section \ref{sub:stack}.
The distance based weighting scheme, $w_d = \sqrt{r_{\perp}} e^{-0.3 r_{\perp}}$, has a maximum weight at 1.67 virial radius of the cluster. This weighting increases the signal by 5~\%, but also increases the averaged noise level similarly.
{The reason why the averaged noise increases, is that when performing a weighted average instead of uniform weights, the galaxies that are suppressed will contribute less to the cancellation of noise.}
Other weighting schemes for distance are not considered, but the studied one does not improve the SNR according to our analysis. Even if the SNR does not improve, a benefit of this method is that one does not need to put a sharp distance cutoff by hand, but rather tune the slope of the weighting function. }

{We also use a mass based weighting scheme, with $w_m = \log_{10}(M_\mathrm{vir}/ 10^{10} M_\odot)$. This weighting increases the signal-to-noise significantly by weighting massive galaxies more than light galaxies. The expected detection of the slingshot signal with $10^{-6}$ noise level increases from 3 $\sigma$ to 5 $\sigma$, and the chance of a false positive decreases from 2.89~\% to 0.02~\%. This suggests that a mass based weighting scheme should be considered when using real observations.
The final weighting scheme we test is a combination of the distance and mass based weighting scheme, but it does not improve the results over the pure mass based weighting scheme. {A possible method for defining a more optimal weighting scheme in a future analysis is via matched filtering.}
For the noise levels of current surveys ($4\times 10^{-6}$), the mass weighting does not increase the SNR sufficiently to avoid the confusion with false positives.}

\subsection{Applications to Modified Gravity}
When using the method discussed in this paper, one estimates an average $m v_x/ r_\mathrm{vir}$ of infalling galaxies around a cluster. This can be combined with other observables, like the velocity along the line of sight and the inferred halo mass from lensing. Comparing with such additional data, the slingshot effect can be used as an independent probe of modified gravity. Many scalar--tensor theories will increase the clustering on scales of kiloparsecs to megaparsecs. For instance, the Chameleon model studied by \cite{Brax2013SystematicModels} shows increased clustering. If the modifications apply on galaxy scales, each galaxy can be more massive and more dense. If the modifications apply on megaparsec scales, galaxies will fall faster due to the fifth forces on large scales \citep{Ivarsen2016DistinguishingStatistics}. Both of these effects would increase the expected slingshot signal with respect to a similar scenario in $\Lambda$CDM, making it a possible probe for enhanced gravity over several different scales. {Because the method described in this paper only measures the combination $m v_x/ r_\mathrm{vir}$, the separate effects of higher infall velocity and higher galaxy density are degenerate.}

{From the halo mass function of the pure Symmetron case in \cite{Hagala2016CosmologicalFields}, we find that in a typical Symmetron scenario, the individual galaxies will have a 20~\% increase in mass. Assuming a {10\%} increase in infall velocity\footnote{{For all Symmetron models except the one with the weakest coupling, \cite{Ivarsen2016DistinguishingStatistics} found a $> 10 \%$ increase in pairwise velocities}}, we estimate that the average $m v_x/ r_\mathrm{vir}$ at redshift zero can increase with about 30~\% {relative to GR}. In the case of uniform weighting of the galaxy maps and $10^{-6}$ noise level, this increase is equivalent to the one standard deviation measurement error on the slingshot signal.}

{We find that the average $m v_x/ r_\mathrm{vir}$ amongst the 4700 most massive clusters in the simulated catalogue has a standard deviation similar to the average of $m v_x/ r_\mathrm{vir}$. This means that, even in the context of $\Lambda$CDM, the amplitude of the slingshot signal around a single cluster is not decided by the mass of the cluster alone, but also by a combination of the surrounding large scale structure and the merger history of the cluster.    
Unless we know such specifics about the studied cluster, we would need to observe $~36$ clusters to reduce the error of the mean for the "universal" $m v_x/ r_\mathrm{vir}$ with a factor $1/\sqrt{36}\approx 1/6$. If we can do this, the uncertainty of our measurements can be low enough for us to begin distinguishing gravitational models like the Symmetron from pure $\Lambda$CDM, with two sigma significance. Both better knowledge of the mass distribution of the surroundings of the cluster, as well as a better weighting scheme for the stacking of galaxy maps, can reduce the amount of clusters needed to distinguish between different gravitational models.}
%





\section{Conclusion}
In this paper, we present a method to detect transverse motions of galaxies by stacking the dipole signal of the slingshot effect. For this method to work, we need to be able to identify a preferential direction to align the galaxies along their expected direction of motion.  This can be done by taking into account galaxies that either fall into clusters or move away from the centre of the voids.  We show a detection strategy for galaxies falling into a nearby cluster, like the Coma Cluster. A similar analysis can be done for galaxies around voids.

The possibility to detect the signal with certainty with CMB-S4 experiments is very high. There are some simplifications done in this paper that should be considered more thoroughly when analysing real data. The most important considerations relate to the choice of cutoff in halo mass, and the cutoff in distance from the central cluster. 
We use a mass cutoff of $M > 10^{11} M_\odot$ when considering a halo for stacking. A too low cutoff means adding mostly noise for each image, while a too high mass cutoff gives fewer halos to stack. Furthermore, we do not include an upper mass limit. This means that we are in practice stacking the dark matter halos of some smaller galaxy clusters as well as individual galaxies. Choosing the halo mass cutoffs in a more sophisticated way---{like using a weighting scheme}---could improve the signal. Increasing the distance within which to consider infalling galaxies will allow for including more galaxies and could result in better statistics. An inner radius cut-off can also be considered, since galaxies within approximately one virial radius of the cluster do not appear to have a preferred radial direction. When excluding these galaxies, a better signal can be expected. {Another option is to use a distance based weighting scheme instead of a hard cut-off. We test a simple distance based weighting scheme, which does not impact the signal-to-noise ratio significantly.  We also test a mass based weighting scheme, which we find to increase the signal-to-noise by weighting massive galaxies more than light galaxies. When used on real data, this weighting scheme can be more or less efficient depending on the confidence of the mass estimates in the galaxy catalogue.}

In this paper, 
we stack the CMB maps centred on $\mathcal{O} \left( 10^4 \right)$ simulated galaxies, and orient them according to their expected infall direction towards a nearby massive cluster. By fitting a dipole template to the stacked signal, we show that the slingshot effect is statistically distinguishable from noise when using the next generation of CMB experiments. By measuring the slingshot signal around $~36$ clusters, we can constrain the signal sufficiently to test alternative theories for gravity.

\section*{Acknowledgements}
Thanks for ideas and comments from several people at the Institute of Theoretical Astrophysics in Oslo, especially: Håkon Dahle, Hans Kristian Eriksen, Marta Bruno Silva, Håvard Tveit Ihle, Øyvind Christiansen, and Daniel Heinesen. A great thanks to Sigurd Kirkevold Næss for helping with technical details.
Many thanks for detailed and helpful comments and questions from the referee, which helped strengthening the paper significantly.
We thank the Research Council of Norway for their support. 
This paper is based upon work from the COST action CA15117 (CANTATA), supported by COST (European Cooperation in Science and Technology).
The CosmoSim database used in this paper is a service by the Leibniz-Institute for Astrophysics Potsdam (AIP).
The MultiDark database was developed in cooperation with the Spanish MultiDark Consolider Project CSD2009-00064.

\bibliographystyle{arxiv}
\bibliography{references}

\onecolumn{}

\section*{Appendix: Calculation of slingshot effect from a spherical halo model}

We have that 
\begin{equation}
\frac{\Delta T_{\mathrm{slingshot}}}{T}=2v_{x}\int\frac{\partial\Phi}{\partial x}\,\mathrm{d}z.
\end{equation}
\\
Since gravitational potentials are additive, we will have a contribution
from the NFW dark matter profile and from the Hernquist profile.

\begin{equation}
\frac{\Delta T_{\mathrm{slingshot}}}{T}=2v_{x}\int\frac{\partial\Phi_{\mathrm{NFW}}}{\partial x}+\frac{\partial\Phi_{\mathrm{Hernq}}}{\partial x}\,\mathrm{d}z.
\end{equation}
 We will now calculate these integrals separately. In principle the
integral limits is from the surface of last scattering and until today,
but as long as the kernel we are integrating peaks around $z=0$,
we can safely integrate from $z=-\infty$ to $z=\infty$ instead.

\subsection*{NFW}

From \cite{okas2001PropertiesProfile}, the gravitational potential of the NFW halo is given by

\begin{equation}
\Phi_{\mathrm{NFW}}=-Gmg\times\frac{\ln\left(1+\frac{c_\mathrm{NFW} r}{r_\mathrm{vir}}\right)}{r}
\end{equation}
\\
where $c_\mathrm{NFW}$ is the concentration (we assume $c_\mathrm{NFW}=15$) and 
\begin{equation}\label{eq:g}
g\equiv\frac{1}{\ln(c_\mathrm{NFW}+1)-\frac{c_\mathrm{NFW}}{c_\mathrm{NFW}+1}}.
\end{equation}

Substituting $r=\sqrt{x^{2}+y^{2}+z^{2}}$, we can do the derivative
with respect to $x$,

\begin{align}
\frac{\partial\Phi_{\mathrm{NFW}}}{\partial x} & =\frac{Gmgx}{r^{2}}\left(\frac{\ln\left(1+\frac{c_\mathrm{NFW} r}{r_\mathrm{vir}}\right)}{r}-\frac{c_\mathrm{NFW} /r_\mathrm{vir}}{1+\frac{c_\mathrm{NFW} r}{r_\mathrm{vir}}}\right).
\end{align}

One can find the indefinite integral
\begin{align}
\intop\frac{\partial\Phi_{\mathrm{NFW}}}{\partial x}\mathrm{d}z & =\frac{Gmgx\left( \frac{r_\mathrm{vir}\arctan\left(\frac{c_\mathrm{NFW}z}{\sqrt{c_\mathrm{NFW}^{2}\left(x^{2}+y^{2}\right)-r_\mathrm{vir}^{2}}}\right)} {\sqrt{c_\mathrm{NFW}^{2}\left(x^{2}+y^{2}\right)-r_\mathrm{vir}^{2}}} -\frac{r_\mathrm{vir}\arctan\left(\frac{r_\mathrm{vir} z} {r\sqrt{c_\mathrm{NFW}^{2}\left(x^{2}+y^{2}\right)-r_\mathrm{vir}^{2}}}\right)}{\sqrt{c_\mathrm{NFW}^{2}\left(x^{2}+y^{2}\right)-r_\mathrm{vir}^{2}}} +\frac{z\ln\left(\frac{c_\mathrm{NFW}r}{r_\mathrm{vir}}+1\right)}{r}-\ln\left(r+z\right)\right)}{x^{2}+y^{2}}\\
\end{align}
We are interested in evaluating this integral with limits $z=-\infty$ and $z=\infty$.
We start by finding the limits of the arctangent expressions. We use that $\lim_{x\rightarrow\pm \infty} arctan\left( x\right) = \pm \pi/2$, and find that the first arctangent has the limit
\begin{align}
    \left. \arctan\left(\frac{c_\mathrm{NFW}z}{\sqrt{c_\mathrm{NFW}^{2}\left(x^{2}+y^{2}\right)-r_\mathrm{vir}^{2}}}\right) \right|_{z \rightarrow \infty }
    &= \lim_{Z\rightarrow\infty} \arctan(Z) = \frac{\pi}{2},
\end{align}
and similarly
\begin{align}
    \left. \arctan\left(\frac{c_\mathrm{NFW}z}{\sqrt{c_\mathrm{NFW}^{2}\left(x^{2}+y^{2}\right)-r_\mathrm{vir}^{2}}}\right) \right|_{z \rightarrow - \infty }
    &= - \frac{\pi}{2}.
\end{align}
To evaluate the limits of the second arctangent, we note that
\begin{equation}
    \lim_{z\rightarrow \pm \infty} \frac{z}{r} = \pm 1.
\end{equation}
This leaves us with the following arguments for the second arctangent:
\begin{align}
    \left.  \frac{r_\mathrm{vir} z}{r \sqrt{c_\mathrm{NFW}^{2}\left(x^{2}+y^{2}\right)-r_\mathrm{vir}^{2}}}\right|_{z \rightarrow \infty }
    &= \frac{r_\mathrm{vir}}{\sqrt{c_\mathrm{NFW}^{2}\left(x^{2}+y^{2}\right)-r_\mathrm{vir}^{2}}}, \\
    \left.  \frac{r_\mathrm{vir} z}{r \sqrt{c_\mathrm{NFW}^{2}\left(x^{2}+y^{2}\right)-r_\mathrm{vir}^{2}}}\right|_{z \rightarrow -\infty }
    &= -\frac{r_\mathrm{vir}}{\sqrt{c_\mathrm{NFW}^{2}\left(x^{2}+y^{2}\right)-r_\mathrm{vir}^{2}}}.
\end{align}

Regarding the logarithmic expressions, the limit at $z\rightarrow\infty$ is
\begin{equation}
    \lim_{z\rightarrow \infty} \frac{z\ln\left(\frac{c_\mathrm{NFW}r}{r_\mathrm{vir}}+1\right)}{r}-\ln\left(r+z\right)
    =\lim_{z\rightarrow \infty} \ln\left(\frac{c_\mathrm{NFW}r}{r_\mathrm{vir}}+1\right)-\ln\left(z+z\right)
    =\lim_{z\rightarrow \infty} \ln\left(\frac{{c_\mathrm{NFW}z/r_\mathrm{vir}}}{2z}\right)
    = \ln\left(\frac{c_\mathrm{NFW}}{2r_\mathrm{vir}}\right).
\end{equation}
The limit of the logarithmic terms when $z\rightarrow -\infty$ is
\begin{align}
    \lim_{z\rightarrow -\infty} \frac{z\ln\left(\frac{c_\mathrm{NFW}r}{r_\mathrm{vir}}+1\right)}{r}-\ln\left(r+z\right)
    &=\lim_{z\rightarrow -\infty} -\ln\left(\frac{c_\mathrm{NFW}z}{r_\mathrm{vir}}\right)-\ln\left(z\sqrt{1 + \frac{x^2 + y^2}{z^2}}-z\right)
    =\lim_{z\rightarrow -\infty} -\ln\left(\frac{c_\mathrm{NFW}z}{r_\mathrm{vir}}\right)-\ln\left(\frac{x^2 + y^2}{2z}\right) \\
    &= -\ln\left(\frac{c_\mathrm{NFW}\left(x^2 + y^2\right)}{2r_\mathrm{vir}}\right),
\end{align}
where we used that $\sqrt{1+x}\approx 1 + x/2$ for small x.

Combining all of those, we are left with
\begin{align}
\intop_{-\infty}^{\infty}\frac{\partial\Phi_{\mathrm{NFW}}}{\partial x}\mathrm{d}z & =\frac{Gmgx}{x^{2}+y^{2}}
\left[
S\left(\pi-2\arctan\left(S\right)\right)+\ln\left(\frac{c_\mathrm{NFW}^{2}\left(x^{2}+y^{2}\right)}{4r_\mathrm{vir}^{2}}\right)
\right]
.
\end{align}
Here, we have defined
 \begin{equation}
     S \equiv \frac{r_\mathrm{vir}}{\sqrt{c_\mathrm{NFW}^{2}\left(x^{2}+y^{2}\right)-r_\mathrm{vir}^{2}}}.
 \end{equation}


\subsection*{Hernquist}

The gravitational potential of a Hernquist halo with mass $m$ is
simply given by
\begin{equation}
\Phi_{\mathrm{Hernq}}=-\frac{Gm}{r+a},
\end{equation}
\\
where $a$ is a scale length, which is related to the half-mass-radius
as $a=\frac{r_{1/2}}{1+\sqrt{2}}$. We chose $r_{1/2}=0.015 r_\mathrm{vir}$
based on figure 1 from \cite{Kravtsov2013TheGalaxies}, where the data indicates $r_{1/2} \approx 0.015 r_{200c}$.

The derivative with respect to the $x$ coordinate is
\begin{equation}
\frac{\partial\Phi_{\mathrm{Hernq}}}{\partial x}=\frac{Gmx}{r\left(a+r\right)^{2}},
\end{equation}
which results in the following indefinite integral along $z$:
\begin{equation}
\intop\frac{\partial\Phi_{\mathrm{Hernq}}}{\partial x}\mathrm{d}z=
\frac{Gmx}{a^2}\left(
\frac{a^2 z}{\left(x^2+y^2-a^2\right) \left(r+a\right)}
+\frac{a^3 \arctan\left(\frac{a z}{r\sqrt{x^2+y^2-a^2}}\right)}{\left(x^2+y^2-a^2\right)^{3/2}}
-\frac{a^3 \arctan\left(\frac{z}{\sqrt{x^2+y^2-a^2}}\right)}{\left(x^2+y^2-a^2\right)^{3/2}}
\right).
\end{equation}
We chose to define 
\begin{equation}
    U \equiv \frac{a}{ \sqrt{x^{2}+y^{2}-a^{2}}},
\end{equation}
which does not depend on $z$.

The limits of the first term when $z\rightarrow \pm\infty$ are
\begin{equation}
\lim_{z\rightarrow\infty} \frac{U^2 z}{ \left(r+a\right)} = {U^2},
\end{equation}
and
\begin{equation}
\lim_{z\rightarrow - \infty} \frac{U^2 z}{\left(r+a\right)} = -{U^2}.
\end{equation}
The limits of the second term are
\begin{equation}
\lim_{z\rightarrow\infty} U^3 \arctan\left(U\frac{z}{r}\right)= U^3 \arctan\left(U\right),
\end{equation}
and
\begin{equation}
\lim_{z\rightarrow-\infty} U^3 \arctan\left(U\frac{z}{r}\right)= - U^3 \arctan\left(U\right).
\end{equation}
The last arctangent converges to $\pm \pi/2$, giving 
\begin{equation}
\lim_{z\rightarrow\infty} -U^3 \arctan\left(U\frac{z}{a}\right) = -\frac{\pi U^3}{2 },
\end{equation}
and
\begin{equation}
\lim_{z\rightarrow-\infty} - U^3 \arctan\left(U\frac{z}{a}\right) = \frac{\pi U^3}{2},
\end{equation}

Finally, the slingshot integral for the Hernquist distribution can be written
\begin{equation}
\intop_{-\infty}^{\infty}\frac{\partial\Phi_{\mathrm{Hernq}}}{\partial x}\mathrm{d}z=Gmx \frac{U^2}{a^2}
\left(
2
+U\left[
2\arctan\left(U\right)
-\pi
\right]
\right).
\end{equation}

\subsection*{Sum}

Because we are assuming that the total mass $m_{DM}$ of the dark
matter halo is in the NFW component, with an additional $m_{DM}/10$
in baryons, we write the combined effect as
\begin{equation}
\frac{\Delta T_{\mathrm{slingshot}}}{T}= 
2v_{x} \intop\frac{\partial}{\partial x}\Phi_{\mathrm{NFW}} \left( m = m_{DM} \right) +
\frac{\partial}{\partial x}\Phi_{\mathrm{Hernq}} \left( m = \frac{m_{DM}}{10} \right)\,\mathrm{d}z
= \frac{2 Gm_{DM} v_{x}}{r_\mathrm{vir}}\left(Q_{\mathrm{NFW}}+\frac{1}{10}Q_{\mathrm{Hernq}}\right).
\end{equation}
Here, we use the following notation for dimensionless coordinates:  $x_r \equiv x/r_\mathrm{vir}$ and $x_a \equiv x/a$. Furthermore,
\begin{equation}
Q_{\mathrm{NFW}}\equiv\frac{gx_r}{x_r^{2}+y_r^{2}}
\left[
\ln\left(\frac{c_\mathrm{NFW}^{2}\left(x_r^{2}+y_r^{2}\right)}{4}\right)
-S\left(2\arctan\left(S\right) - \pi\right)
\right],
\end{equation}
and
\begin{equation}
Q_{\mathrm{Hernq}}\equiv\left( \frac{1 + \sqrt{2}}{0.015} \right) \frac{x_a}{x_a^2 + y_a^2 - 1}
\left[
2
+U\left(
2\arctan\left(U\right)
-\pi
\right)
\right].
\end{equation}
The factor of $\left(1 + \sqrt{2} \right) /0.015$ is to convert from the Hernquist scale $a$ to $r_\mathrm{vir}$. We repeat the definitions
 \begin{equation}
     S \equiv \frac{1}{\sqrt{c_\mathrm{NFW}^{2}\left(x_r^{2}+y_r^{2}\right)-1}},
 \end{equation}
\begin{equation}
    U \equiv \frac{1}{ \sqrt{x_a^{2}+y_a^{2}-1}}.
\end{equation}


Note that $S$ and $U$ can become imaginary for light passing close to the centre of the halo. Specifically, both $U$ and $S$ are $\in \left[ -i, -\infty i \right)$. However, $z\left(2\times\arctan\left(z\right) - \pi\right)$ always has one real value, even for imaginary $z$.

Proof: For a real $x > 1$, it follows that $z=-ix$ is negative imaginary with the same domain as $U$ and $S$. Using the logarithm definition of the complex arctangent, we have
\begin{align}
&\left(-ix\right)\left(2\times\arctan\left(-ix\right) - \pi\right)=
\left(-ix\right)\left(2\times\frac{i}{2}\ln\left(\frac{1 - x}{1 + x}\right) - \pi\right) \\
\mathrel{\mathop{=}\limits_{\hphantom{\mathrm{choice}}}} &
x\left(\ln\left(\frac{1 - x}{1 + x}\right) + \pi i \right) =
x\left(\ln\left(\frac{x - 1}{1 + x}\right) + 2 \pi i \right) \\
\mathrel{\mathop{=}\limits_{\mathrm{choice}}} &
x \ln\left(\frac{x - 1}{1 + x}\right).
\end{align}
In the last line, we used the fact that for complex logarithms, $\ln \left( z \right) = \ln \left( - z \right) + \pi i$. Furthermore, any addition of $ 2 \pi i$ can be cancelled by the corresponding free choice of $2 k \pi i$ in the multi-valued complex logarithm. With $x > 1$, this result proves that there is always a real branch of the expression $z\left(2\times\arctan\left(z\right) - \pi\right)$.
This expression is also continuous for values of $r^2 = x^2 + y^2$ crossing through the singularity in $U$ or $S$.
\end{document}